# *Spin Filter Properties of Armchair Graphene Nanoribbons with Substitutional Fe Atoms*

Frank Hagelberg[a], Alexander Kaiser[b], Ivan Sukuba[b], Michael Probst[b]

[a]Department of Physics and Astronomy, East Tennessee State University,

Johnson City, TN 37614, USA

[b]Institut fuer Ionenphysik und Angewandte Physik, University of Innsbruck, Technikerstr. 25,

6020 Innsbruck, Austria

**Abstract:** The spin filter capability of a (0,8) armchair graphene nanoribbon with Fe atoms at substitutional sites is investigated by density functional theory in combination with the non-equilibrium Green's function technique. For specific arrangements, a high degree of spin polarization is achieved. These include a single substitution at an edge position or double substitution in the central sector of the transmission element. The possibility of switching between majority and minority spin polarization by changing the double substitution geometry is predicted. Including the bias dependence of the transmission function proves to be essential for correct representation of the spin-resolved current-voltage profiles.

**Keywords:** Spin transport, Graphene nanoribbons, Non-equilibrium Green's function, Spintronics, Spin filter

Word count: 4,499

## I. Introduction

The question of how to implement basic devices of spintronics circuitry at the nanoscale has attracted much interest [1]. These devices include spin valves [2,3], spin transistors [4,5], and spin filters [6,7]. A common feature of these designs is that they subject charge carriers to magnetic interactions in order to generate or manipulate spin currents. Both theoretical and experimental studies have identified graphene as a highly suitable medium for spin currents which is rationalized by its extreme mechanical and thermal stability [8,9], its high charge carrier mobility [10], and, most importantly, its long spin relaxation times [11,12]. Naturally, research on transmission elements for spintronics applications has focused on graphene nanoribbons, periodic graphene bands of finite width which are obtained by dimensional reduction of the graphene sheet.

Several computational studies have focused on the spin filtering effect of zigzag graphene nanoribbons (zGNRs) [13-15]. These systems exhibit a magnetic ground state [16-18]. Specifically, their magnetic structure may be characterized as a mixture of ferromagnetic (FM) and antiferromagnetic (AFM) order. While FM coupling is found along both zGNR edges, the magnetic moments of these edges are anti-parallel with respect to each other, corresponding to AFM correlation. As has been recognized early, transmission elements based on zGNRs act as spin filters when the FM order of one of these edges is broken, such that the zGNR as a whole adopts non-zero spin polarization [19].
Yan et al. used the non-equilibrium Green's function (NEGF) formalism in conjunction with density functional theory (DFT) to investigate zGNRs with Fe atom impurities, where Fe atoms substitute for some of the terminating H atoms at the nanoribbon edges [13]. In particular, up to four Fe impurity atoms attached to the edges of a narrowed zGNR junction were included in this simulation. The presence of the transition metal component turned out to have a substantial impact on the quantum transport properties of the zGNR transmission element. Thus, a sizable effect on the resistance of the nanoribbon was noted, giving rise, in some cases, to negative differential resistance (NDR). Further, complex spin transport phenomena were found to be induced by the Fe moiety, as documented by spin-resolved current-voltage profiles that were seen to depend sensitively on the geometric arrangements of the adsorbing Fe atoms. Most importantly, for selected configurations the spin polarization degrees of the current traversing the zGNR were shown to reach 80 -100 % in a bias window of about 0.5 to 1.0 V.

A recent study extended this research to silicene nanoribbons, employing Fe, Co, and Ni as transmission metal impurities, and including the relative orientation of their magnetic moments as an additional degree of freedom [15].

Jaiswal *et al.* [20] performed DFT calculations involving periodically repeated Ni atoms in armchair graphene nanoribbons (aGNRs) of four to nine C atoms in width. An analysis of the band structure as well as the partial-density-of-states (PDOS) distributions led to the prediction of current spin polarization effects in these qualitatively different type of GNR. Both terminating and substitutional impurity sites were considered. As the respective PDOS distributions revealed significant differences between the spin-up and the spin-down moiety close to the Fermi edge for Ni atoms in terminating and single-edge substitutional positions, the authors concluded that aGNRs with Ni atom impurites might have substantial spin filtering effects in selected geometric arrangements.

Jaiswal and Srivastava [21] investigated the transport properties of aGNRs with substituting or terminating Fe-atoms where the latter form atomic cables that connect the electrodes. Comparing the density of states (DOS) distributions for both spin orientations, the authors obtained degrees of spin polarizations of up to 60% for these metallic materials. Further, the spin polarization magnitude was shown to exhibit a sensitive dependence on the site of the Fe line defect. An analogous study on Fe-doped zGNRs yielded spin polarization degrees up to 95 % [22]. It is noteworthy that Fe atoms at substitutional sites are non-magnetic in graphene, while they adopt a finite magnetic moment in aGNRs [23].

The present work explores the impact of Fe atoms at selected substitutional sites on the transport characteristics of aGNRs. As the aGNR ground state does not display spin polarization at the edges of these structures, they provide a non-magnetic reference medium, such that any spin-dependent effects must be attributed to the presence of the Fe impurities. Specifically, we will determine the spin transport properties of finite aGNRs with one or two substitutional Fe atoms, and we will address the question of optimizing the spin-filtering efficiency of these composites.

## II. Methods

The electronic structure of the systems investigated in this work is obtained by DFT calculations at the level of the local spin density approximation (LSDA) in conjunction with norm-conserving pseudopotentials of the Troullier-Martins type [24]. In particular, the Perdew-Zunger exchange-correlation functional [25] has been used.

Geometry optimizations were carried out with an atomic force threshold of 0.05 eV/Å as nuclear convergence criterion, while the electronic steps were constrained to converge with an accuray of $1.0*10^{-6}$ eV. The basis set size was restricted by a cutoff energy of 2041 eV (= 75 Hartree), and the Monkhorst-Pack method [26] was applied, with a wave number mesh of dimension 1 × 1 × 100. A double-ζ basis set was adopted to represent the electronic states.

Our transport computations are based on the NEGF procedure as implemented in the code Atomistix ToolKit (ATK) [27-30]. In particular, the spin-dependent transmission probability $T_\sigma$ was evaluated as a function of energy by using the relation

$$T_\sigma(E, V_{bias}) = \mathrm{Tr}\{\mathbf{\Gamma_L G \Gamma_R G^\dagger}\} \qquad (I)$$

which contains the voltage across the transmission element, $V_{bias}$, as a parameter. The function $T_\sigma$ is defined as the trace over the matrix $\mathbf{\Gamma_L G \Gamma_R G^\dagger} = \mathbf{t\, t^\dagger}$ where the symbols $\mathbf{\Gamma_i}$ (i = L, R) denote the anti-Hermitian components of the self-energy for the left (L) and the right (R) contact, while **G** stands for the energy dependent matrix of the Green's function for the transmission element. Further, **t** refers to the *transmission matrix* whose elements $t_{nm}$ are amplitudes for the transition of an electron from state *n* of one of the two leads into state *m* of the other [31]. To obtain the transmission probability at $V_{bias} \neq 0$, the effective electrostatic potential induced by the electrodes across the transmission element was calculated by use of the Poisson equation.

The spin-dependent current at a given voltage $V_{bias}$ is obtained from [32]:

$$I_\sigma(V_{bias}) = e/h \int T_\sigma(E, V_{bias}) [f_L(E - \mu_L) - f_R(E - \mu_R)]\, dE. \qquad (II)$$

Here, $f_i(E - \mu_i)$ (i = L, R) denote Fermi-Dirac distribution functions. The expressions $\mu_i$ (i = L, R) are defined as the electrochemical potentials of the left and the right electrode. They are functions of the voltage $V_{bias}$. Specifically:

$$\mu_{L/R}(V_{bias}) = \mu_i(0) -/+ eV_{bias}/2 \qquad (III)$$

The structure of our (0,8) model aGNR with a width of 8 C atoms is as follows: a central region (the *transmission element,* see Figure 1(a)) is placed between semi-infinite aGNR electrodes which are modeled by use of periodic boundary conditions. For the non-periodic directions, a vacuum spacing of 10 Å was chosen. This system was adopted as a compromise between system size and energy gap. The latter tends to increase with diminishing width [33]. A large aGNR energy gap, however, implies insufficient electric screening qualities and may thus lead to a finite electric field in the electrode regime, unless the length of the transmission element is prohibitively large [34]. In this work, transmission elements of 16 aGNR unit cells in length were used. This length was determined by test calculations involving (0,8) nanoribbons with two substitutional Fe atoms at three different lengths, namely 6, 12, and 16 unit cells. For the latter system, convergence was attained with respect to the magneto-current ratio MCR, defined as

$$MCR = (I\uparrow - I\downarrow)/(I\uparrow + I\downarrow), \qquad (IV)$$

where $I\uparrow(I\downarrow)$ stands for the current with up (down) spin orientation. Calculating MCRs for selected voltages in the interval [0, 1V], we noticed substantial differences between the models with 6 and 12 unit cells, while only small deviations, in the range of one percent or lower, were found when comparing the models with 12 and 16 unit cells.

### III. Results and discusson

We first discuss structures with a single Fe atom impurity at a substitutional site and then turn to two-atom substitution. For both classes of materials, several protoypical arrangements are explored in terms of their spin polarizing efficiencies.

A) *One substitutional Fe atom* (*1Fe-aGNR*).

Panels (b) and (c) of Figure 1 show two configurations that involve a single Fe atom embedded into the (0,8) aGNR reference system (Figure 1(a)). Specifically, we compare a bulk substitution site 1(b) with an edge site 1(c).

Naturally, implanting an Fe-atom impurity leads to a local expansion of the graphene lattice. Specifically, the bond length between the Fe center in 1(b) and the nearest neighbor C atom is found to increase by 19% compared to the C-C bond length in 1(a). An independent plane-wave DFT (PW-DFT) computation [35] where the transmission element of structure 2(b) was simulated by a cluster model, involving hydrogen termination of the zigzag edges, arrived at a margin of 20%. Evaluating the total energies $E$ of structures 1(b) and 1(c), we find that $E$(1(c)) is lower than $E$(1(b)) by 1.46 eV. This finding is readily rationalized as a consequence of the geometric distortion induced in the graphene lattice by the central guest atom, which is less pronounced for edge substitution. The observed trend is in qualitative agreement with computational findings on aGNRs doped or terminated with Fe impurities that form linear substructures between the electrodes [21]. The formation energy of these systems, defined as the energy release upon incorporation of the Fe component, tends to be lowest for center site substitution and highest for Fe-termination of the aGNR.

[Insert Figure 1 near here]

Figure 2 contains a comparison between the total density of states (DOS) profiles of the pristine (0,8) aGNR and the Fe doped systems shown in Figure 1. Going from the pure (0,8) system to the aGNR-Fe structures, we find that the symmetry between the majority and minority spin subsystems is broken. In particular, the Van-Hove singularities closest to the Fermi level in these subsystems are shifted with respect to each other.

[Insert Figure 2 near here]

This effect is related to hybridization of the Fe impurity with the aGNR lattice atoms in its proximity, as illustrated by Figure 3 which shows the PDOS contributions from the Fe atom as well as the three C atoms closest to it (the *nearest neighbor shell*) and the six second closest C atoms (the *next-nearest neighbor shell*). The distributions due to the next-nearest neighbors exhibit sharply defined maxima at E = -0.22 eV and -0.73 eV, i.e. the energies of the PDOS maxima due to the occupied Fe states of highest energy. In contrast, the analogous decomposition for the nearest neighbor shell yields a marked difference between the spin-up and the spin-down distributions, as the former contributes more strongly to the state at E = -0.73. From this observation, the interaction between spin majority states of the Fe atom and the nearest-neighbor shell exceeds that between minority states. This effect, in turn, might be correlated with the shift between the majority and minority PDOS maxima of Fe close to $\varepsilon_F$.

[Insert Figure 3 near here]

The magnetocurrent ratios (MCRs) given in Table 1 suggest that sizable spin-filtering effects can be attained already with a single substitutional Fe atom in the aGNR network. The onset of transmission is found at a minimal bias of 0.2 V which reflects the band gap size $\Delta E = 0.2$ eV of the (0,8) aGNRs considered in this work. A scheme of the electrode band structure in the vicinity of the Fermi level is provided in the Supplemetary Information section.

We note that the spin polarization effects achieved by the 1Fe-aGNR systems included here vary sensitively with the substitutional site and with the bias voltage. Comparing the two configurations in terms of their efficiencies as spin filters, one finds that structure 1(c) performs best, yielding MCRs in excess of 0.5 in a sizable voltage interval. This finding may be discussed in terms of transmission matrix eigenstates. These states were computed for structures 1(b) and 1(c). In both cases, the eigenstate calculation was based on inspecting the transmission spectrum. The energies of the transmission maximum closest to the Fermi level were identified, and the transmission matrix eigenstates were evaluated at these energies, namely $E = -0.22$ eV for structure 1(b), and $E = -0.58$ eV for structure 1(c). Figure 4 shows the obtained scattering states that spread through the transmission elements. For 1(c), the spin-up component is strongly reduced when compared with the spin-down component. In contrast, similar amplitudes are found for both components in case of 1(b). In accordance with this result, the spin filtering performance of structure 1(c) is superior to that of structure 1(b).

[Insert Figure 4 near here]

*B) Two substitutional Fe atoms (2Fe-aGNR)*

Several double substitution geometries were inspected. Figure 5 presents the three 2Fe-aGNR configurations considered in this work.

[Insert Figure 5 near here]

Figure 6 shows the projected density of states (PDOS) distributions for the two Fe atoms of structure 5(a). We distinguish between the configurations with parallel and anti-parallel spins of the Fe atoms. The PDOS distributions shown in Figure 6 refer to the latter case. For the Fe atom on the left (right), the PDOS maximum closest to the Fermi level ($\varepsilon_F$) is contributed by a state with majority, or spin-up (minority, or spin-down) orientation, followed by a maximum due to the minority (majority) orientation.

[Insert Figure 6 near here]

These patterns are reflected by the respective transmission spectra at zero bias, as shown in Figure 7. As the maximum nearest (next-nearest) to the Fermi level in the PDOS distribution for two Fe atoms with parallel spin oriention can be clearly assigned to the spin-down (spin-up) moiety, the profile of the corresponding spin-resolved transmission spectra is plausible (see Figure 7(a)). In particular, the first two transmission maxima belong to opposite spin orientations, and their energy difference equals that between the first two PDOS maxima below $\varepsilon_F$. In the spin-antiparallel case, where both Fe atoms equally contribute to both PDOS maxima, the spin-resolved transmission profiles at zero bias almost coincide (see Figure 7(b)).

[Insert Figure 7 near here]

In accordance with the transmission spectrum shown in Figure 7(a), spin-down transmission is dominant in the bias window [0, 1V]. This is confirmed by the black curve in Figure 8 which once more refers to the case of parallel spins. Specifically, the difference ($I\uparrow - I\downarrow$) is shown for $0 \leq V_{bias} \leq 1.0$ V. The red curve depicts the current difference for the alternative of anti-parallel Fe atom spins. Again, the spin-down current $I\downarrow$ outweighs the spin-up current $I\uparrow$. However, the onset of this effect is delayed, since there is little difference between the two spin-resolved transmission spectra for $V_{bias} \leq 0.2$ V, whereas for $V_{bias} > 0.2$ V, a distinct splitting is observed.

[Insert Figure 8 near here]

The difference between the current-voltage profiles associated with the two spin configurations of the Fe impurity atoms has marked consequences for the spin filtering properties of the two arrangements. From Table 2, the magnetocurrent ratios MCR for the spin-parallel configuration in the voltage interval [0.2 V, 1.0 V] are large, with a maximum magnitude of 0.970, while the MCR results for the spin-antiparallel alternative are consistently lower.

We emphasize that the spin polarization effect in case of antiparallel spins of the two Fe impurities can only be simulated if the change of the transmission function with the voltage, $V_{bias}$, is taken into account.

Turning to structure 5(b), we make qualitatively similar observations. Here, the distance between the two atomic impurities is increased by a factor of 2.2. As in model 5(a), both atoms occupy inequivalent aGNR lattice sites. The enhanced separation leads, in the voltage interval [0.2 V, 0.7 V], to MCR values approximately half as large as those found for structure 5(a). Once more, the minority spin orientation is preferentially transmitted.

In structure 5(c), two Fe atoms are placed at two corners of a carbon hexagon, occupying sites of the same graphene sublattice. This substitutional arrangement may be characterized as a motif consisting of two Fe atoms bridged by a C atom. From the total DOS distribution shown in Figure 9, the first DOS maximum below the Fermi edge is here due to majority spin states (spin-up). This feature is reflected by the corresponding magnetocurrent ratios (see Table 2). Sizable majority spin polarization is obtained in the voltage regime [0.3 V, 0.7 V].

[Insert Figure 9 near here]

For deeper understanding of this observation, we constructed the *molecular projected self-consistent Hamiltonian* (MPSH) for the 2Fe subunit of system 5(c). The MPSH emerges from projecting the self-consistent Hamiltonian of the overall system onto the Hilbert space defined by the basis functions of the selected molecular subgroup. No energy eigenvalue matching the majority spin energy of the DOS maximum closest to the Fermi level, -0.12 eV, could be identified in the resulting molecular spectrum. This finding suggests that the low-energy maximum in the spin-up DOS distribution is not directly caused by the mutual interaction of the Fe atoms. However, repeating this analysis with the $Fe_2C_5$ substructure that is obtained by selecting the two impurity atoms and the five C atoms bonding to them (see Figure 5(c)), we find that the highest occupied molecular orbital (HOMO) eigenvalue for the spin-up subsystem coincides with the energy of the DOS maximum closest to the Fermi of structure 5(c). We conclude that this feature arises from the interaction between the impurity atoms and the π states of the C atoms attaching to them.

## IV. Conclusion

The spin-dependent transmission properties of (0,8) graphene nanoribbons (GNRs) with Fe atoms at substitutional sites have been studied by use of the non-equilibrium Green's function (NEGF) method in conjunction with density functional theory (DFT). Emphasis was placed on the effectiveness of these nanomaterials as spin filters. Several configurations of the type $n$Fe-aGNR, with $n =1, 2$, were analyzed in terms of the magnetocurrent ratio (MCR) as a function of the bias. For both single and double substitution, the results are seen to vary sensitively with the number and the substitutional sites of the atomic impurities. A further parameter of relevance for the case of double substitution is the relative orientation of the magnetic moments of the two impurities.

MCR values exceeding 80 percent were recorded for both single and double substitution. In particular, high spin polarization efficieny was found for an arrangement involving two Fe atoms substituted at bulk sites along the aGNR length coordinate. This effect was seen to diminish with increasing distance between the impuritiy atoms. Substituting two adjacent Fe atoms along the width coordinate turned out to reverse the sign of the spin polarization in a wide interval of voltages. Qualitative explanations of the oberserved spin polarization effects could often be given in terms of the transmission spectrum at zero bias in combination with total and partial density-of-states distributions. For adequate description of the current-voltage profiles, however, taking into account the bias dependence of the transmission function proved to be imperative.

For a more thorough understanding of the findings described here, we plan to extend this study to other graphene-based transmission elements, and to include a wider range of transition metal impurities as well as substitution geometries. Further, it will be interesting to examine some basic assumptions on which the present work relies. Thus, spin relaxation processes have been neglected. These arise primarily from spin-orbit interactions [36,37]. By both computational estimate [11] and experimental observation [12], these effects have been found to be small in graphene. This assessment, however, could change when admission is made for transition metal impurity atoms in a graphene transmission element. Similar considerations hold, in principle, for spin relaxation processes due to hyperfine interactions. The latter are expected to be small in the present case since the nuclear spin of the most abundant Fe isotope, $^{56}$Fe, vanishes in the ground state of the nucleus. More comprehensive future simulations of spin filter effects in GNRs, however, will have to incorporate spin relaxation channels.

# Tables

**Table 1**: Magnetocurrent ratios MCR for structures 1(b) and 1(c) in the voltage regime $V_{bias} \leq$ 1.0 V.

| $V_{bias}$ [V] | MCR (1(b)) | MCR (1(c)) |
|---|---|---|
| 0.2 | 0.38 | -0.622 |
| 0.3 | 0.3 | -0.396 |
| 0.4 | 0.125 | -0.459 |
| 0.5 | -0.054 | -0.518 |
| 0.6 | -0.142 | -0.562 |
| 0.7 | -0.134 | -0.575 |
| 0.8 | -0.131 | -0.578 |
| 0.9 | -0.107 | -0.581 |
| 1 | -0.111 | -0.585 |

**Table 2**: Magnetocurrent ratios for structures 5(a) – (c) in the voltage regime $V_{bias} \leq$ 1.0 V. Structure 5(a): MCR(↑↑) and MCR(↑↓) refer to the magnetocurrent ratios for parallel and anti-parallel orientation of the Fe atom spins. In models 5(b) and 5(c), the spin orientations of the two Fe atoms are parallel to each other.

| $V_{bias}$ | MCR(↑↑) (5(a)) | MCR(↑↓) (5(a)) | MCR(5(b)) | MCR(5(c)) |
|---|---|---|---|---|
| 0.2 | -0.925 | 0.432 | -0.549 | -0.347 |
| 0.3 | -0.95 | 0.393 | -0.44 | 0.712 |
| 0.4 | -0.965 | 0.394 | -0.448 | 0.818 |
| 0.5 | -0.97 | 0.283 | -0.457 | 0.853 |
| 0.6 | -0.938 | -0.17 | -0.427 | 0.773 |
| 0.7 | -0.898 | -0.409 | -0.357 | 0.555 |
| 0.8 | -0.787 | -0.507 | -0.28 | 0.374 |
| 0.9 | -0.722 | -0.549 | -0.23 | 0.323 |
| 1.0 | -0.682 | -0.619 | 0.367 | -0.208 |

# Figures

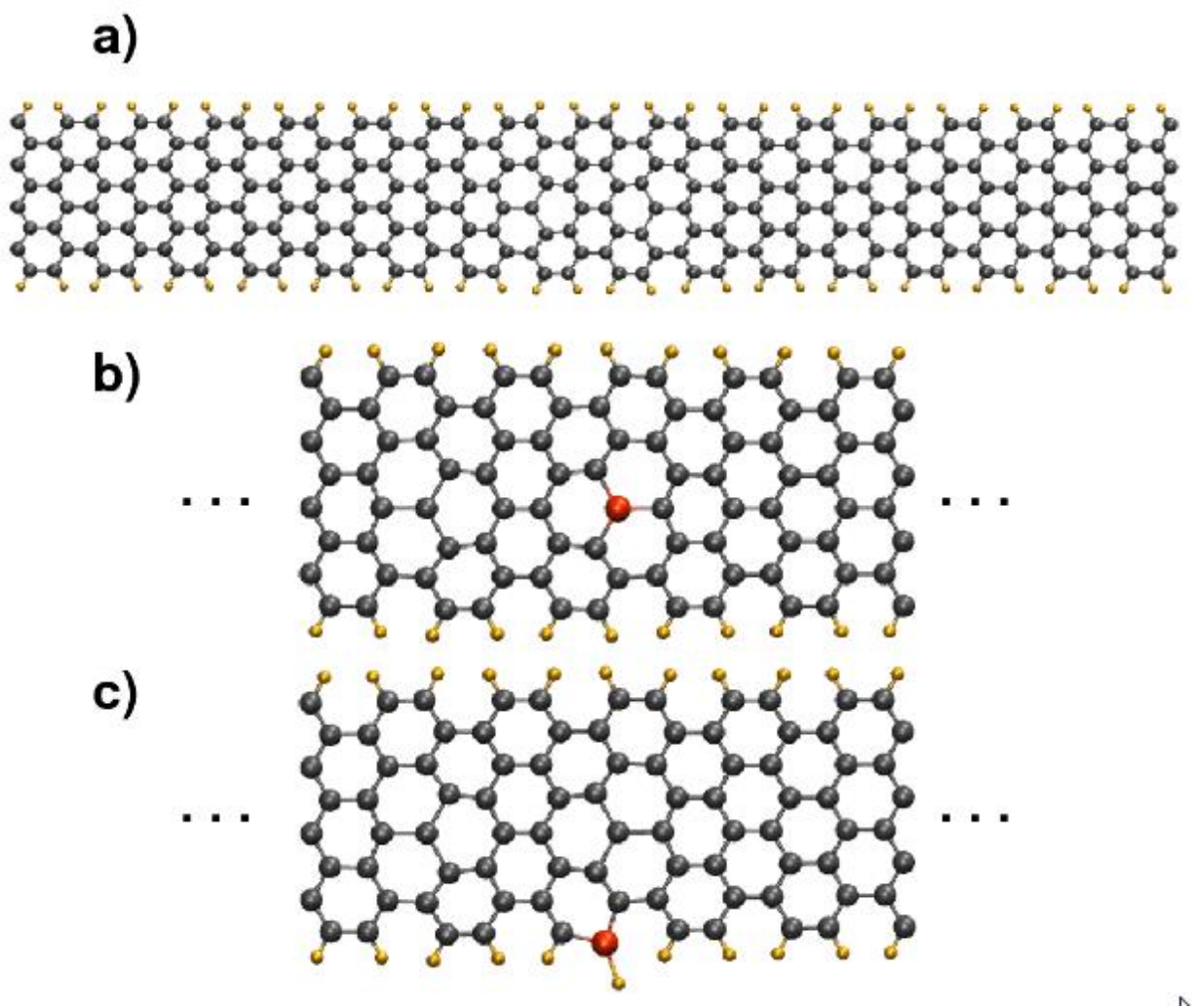

**Figure 1**: The (0,8) transmission element adopted in this work (a) with a single Fe-atom impurity at a center site (b), and at an edge site (c). In all cases, the aGNR transmission element has a length of 16 unit cells and is placed between semi-periodic aGNR electrodes. The full central region is shown in panel (a), while panels (b) and (c) display a segment that contains the Fe impurity.

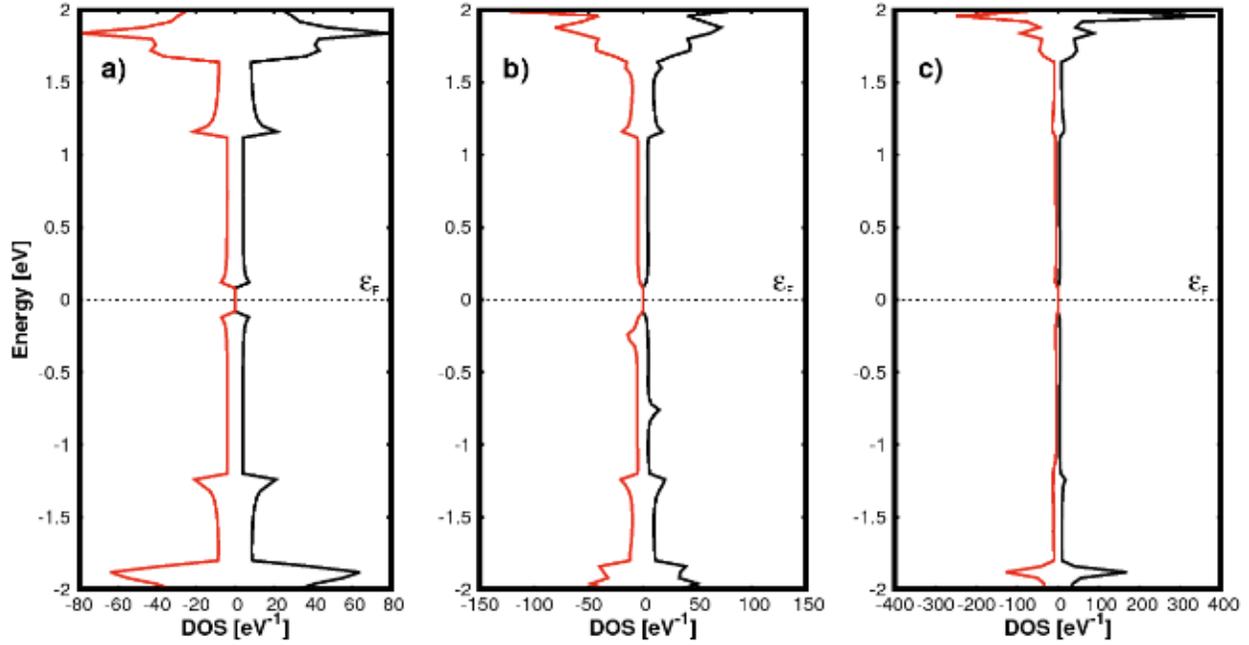

**Figure 2**: Total density of states (DOS) distributions in the vicinity of the Fermi level for the pristine aGNR reference structure (2(a), see Figure 1(a)), and further for the configurations (b) and (c) of Figure 1 (shown in panels 2(b) and 2(c), respectively). The black (red) line represents the spin-up (spin-down) DOS. The Fermi level $\varepsilon_F$ is identified with the zero of energy.

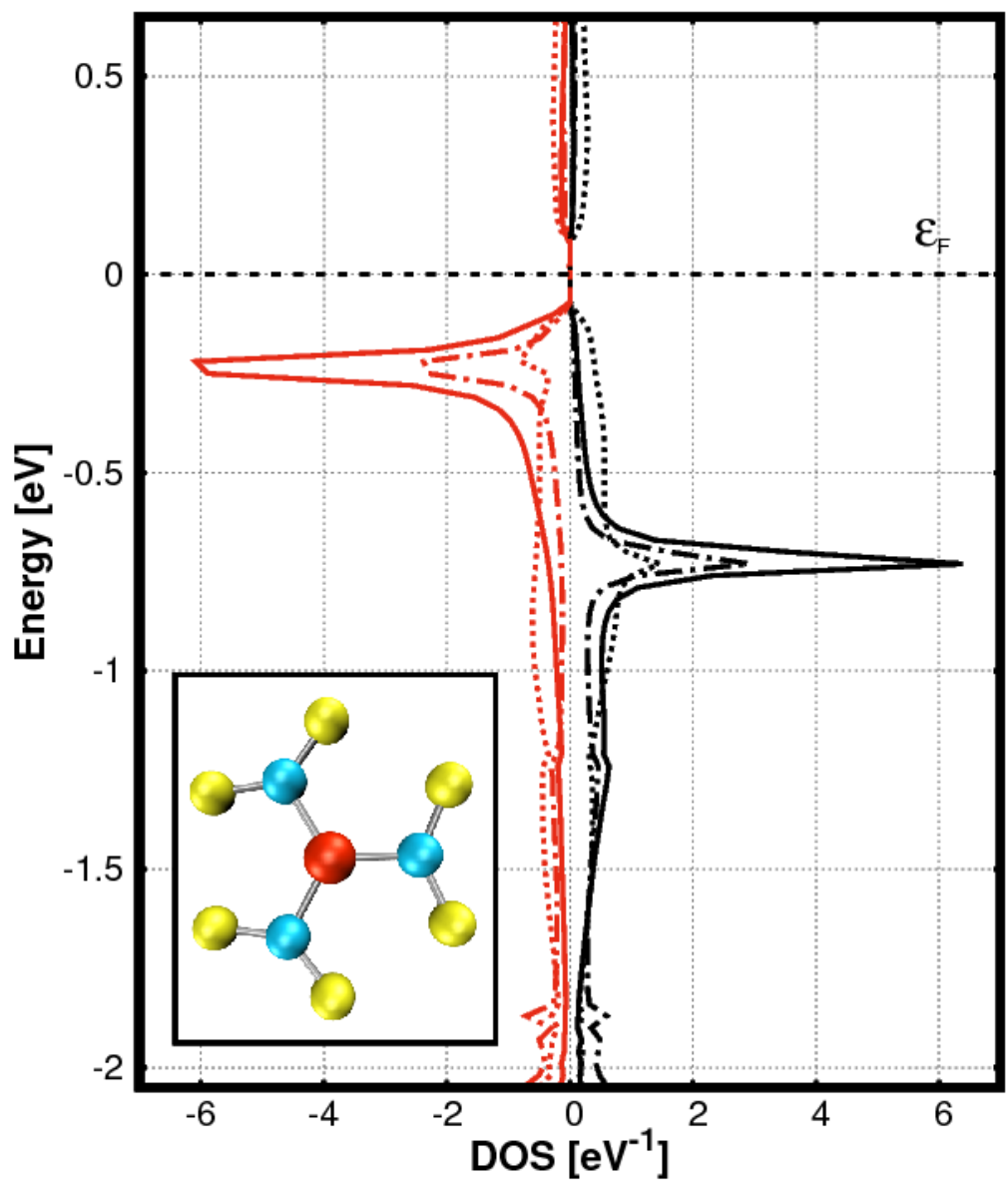

**Figure 3**: Spin-up (black) and Spin-down (red) partial density of states (PDOS) distributions in the vicinity of the Fermi level for geometry 1(b). Included are the Fe impurity states (solid curves) as well as the states due to the nearest-neighbor and the next-nearest neighbor shell of the Fe atom, as indicated by dotted and dot-dashed lines, respectively. The C contributions were normalized to account for the different C atom multiplicities of the two shells.

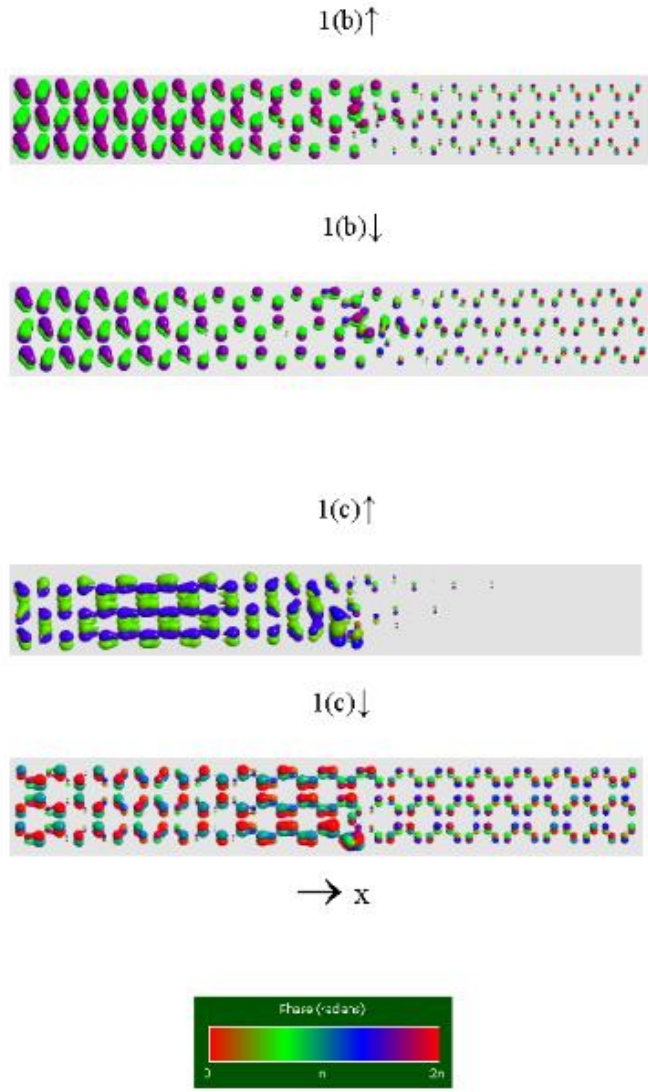

**Figure 4**: Three-dimensional representation of the transmission matrix eigenstates for structures 1(b) and 1(c) at the maxima closest to the Fermi edge in the transmission spectra of the two models. An isovalue of 0.2 was adopted. The eigenstates are complex-valued. While the isovalue indicates the local magnitude of the wave function, the phase factor $e^{i\alpha}$ is encoded by color, with the local phase $\alpha$ defined by the color bar.

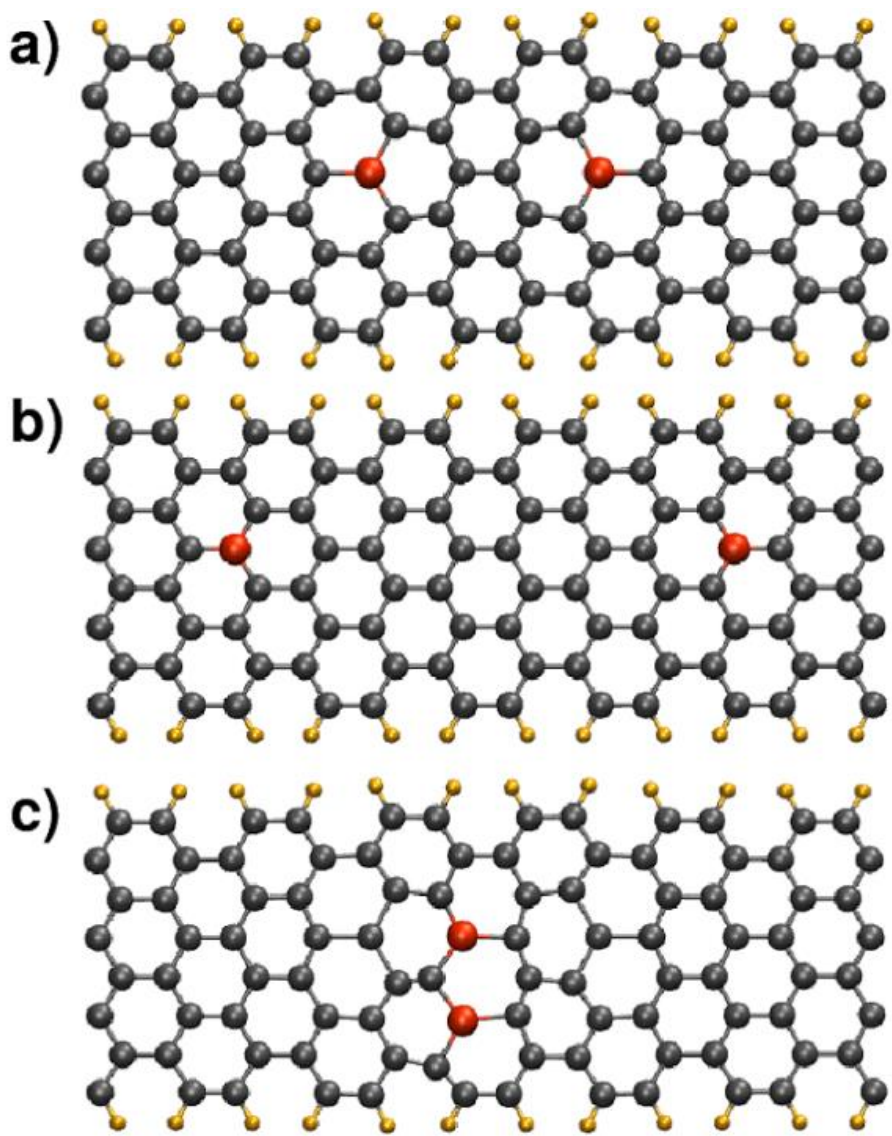

**Figure 5**: Selected test geometries for double substitution: two Fe-atom impurities in a GNR of type (0,8). In all cases, segments of the transmission element that contain the two Fe impurity atoms (red) are shown. The overall length of the transmission element is 16 unit cells (see Figure 1(a)).

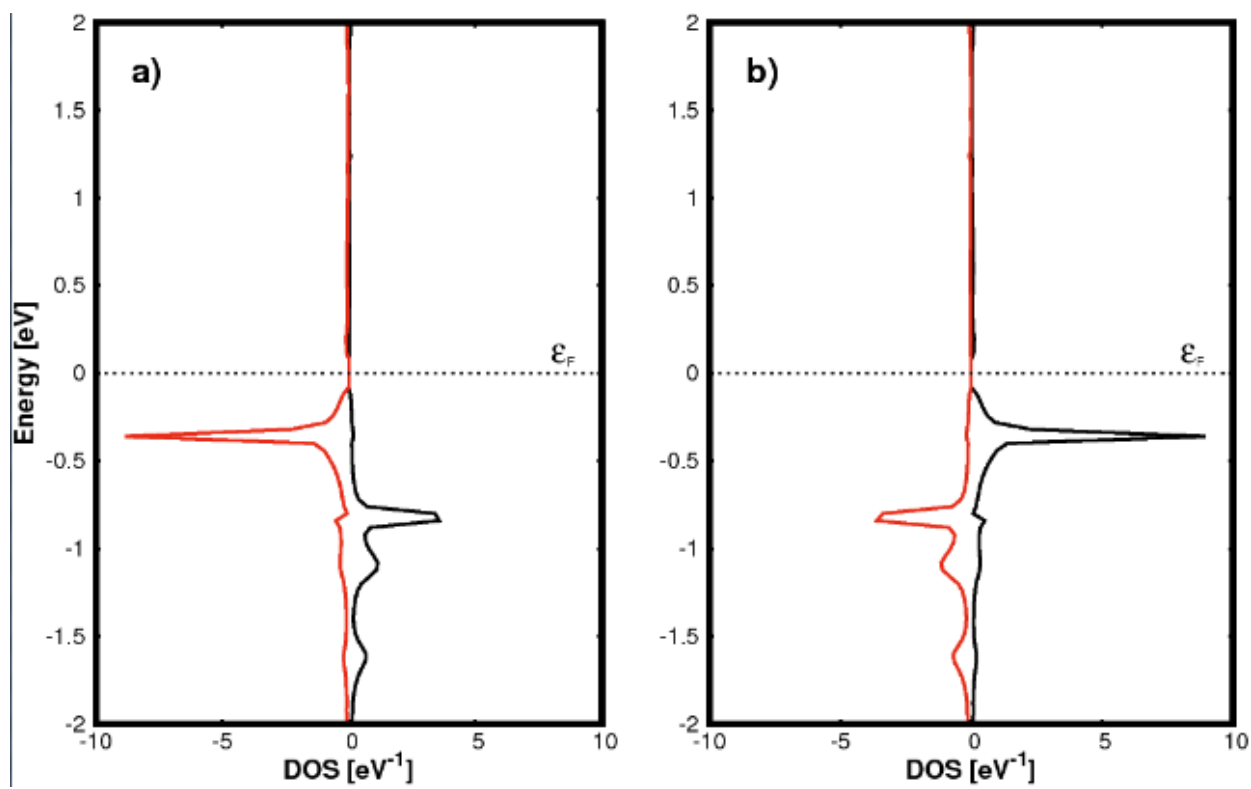

**Figure 6**: Spin-up (black) and spin-down (red) partial density of states (PDOS) distributions for structure 5(a) with anti-parallel spin orientation of the two substitutional atoms. The density of states is projected on the left (a) and the right (b) Fe at atom. The black (red) line represents the spin-up (spin-down) PDOS.

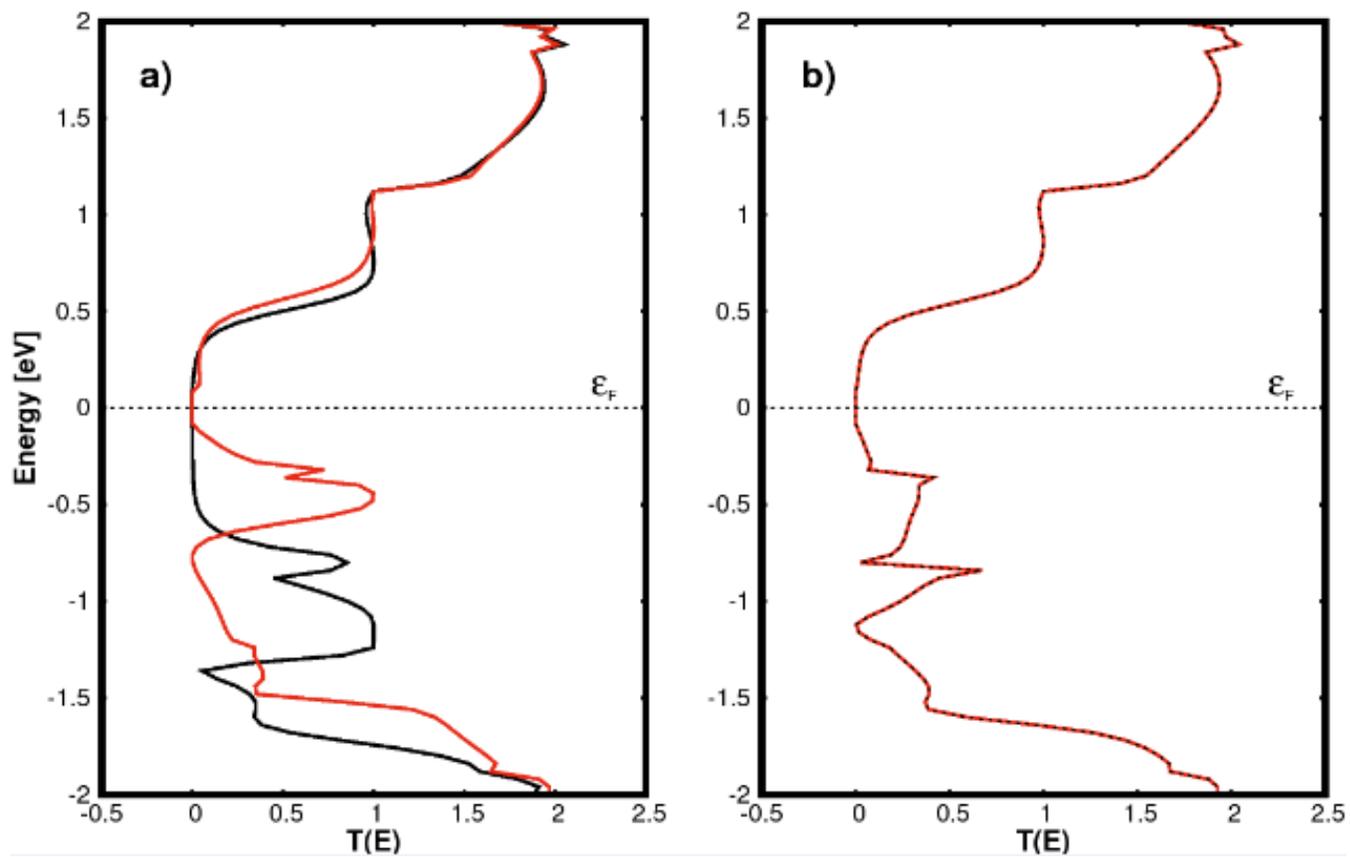

**Figure 7**: Transmission spectra at zero bias for parallel (a) and anti-parallel (b) Fe-atom spin orientations in structure 5(a). The black (red) line refers to the spin-up (spin-down) moiety.

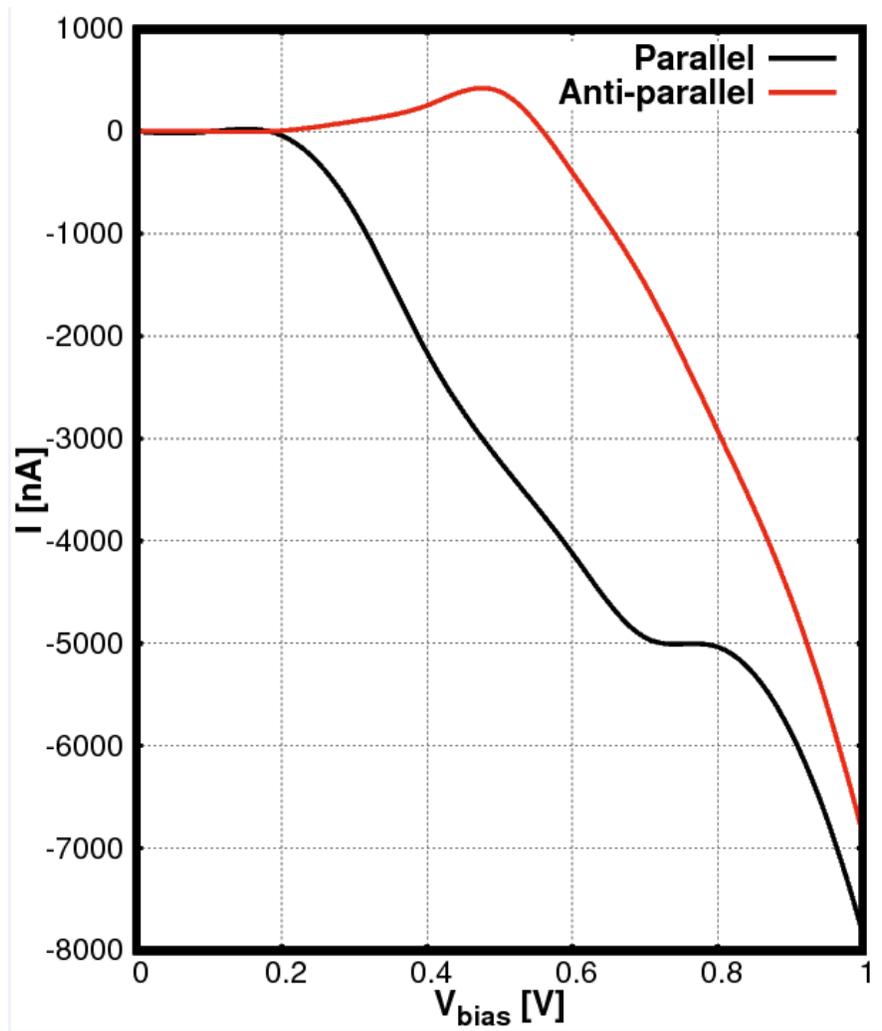

**Figure 8**: Difference between the spin-up and spin-down current (I↑ - I↓) as a function of the voltage $V_{bias}$ for structure 5(a). The black and the red curve refer to parallel and anti-parallel Fe-atom spin orientations, respectively.

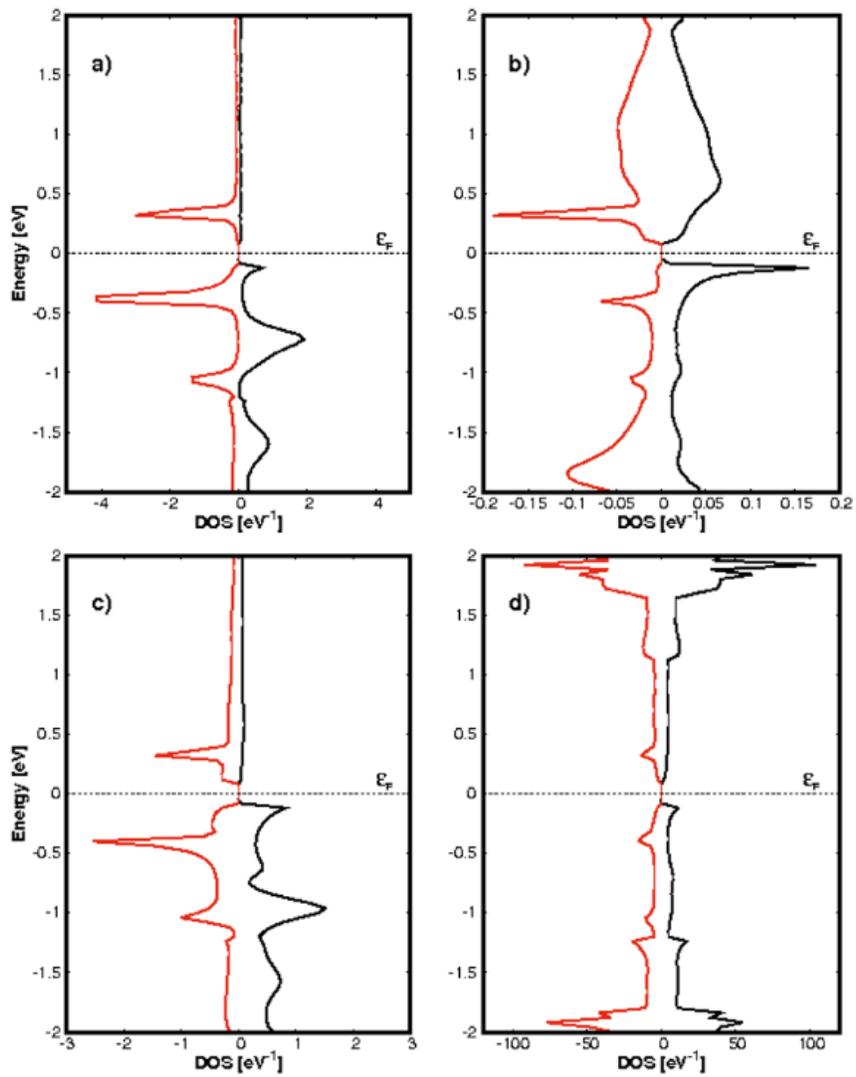

**Figure 9**: Panels (a), (b), and (c) show spin-up (black) and spin-down (red) partial density-of-states (PDOS) distributions for structure 5(c), projected on the lower Fe atom (a), on the C atom bridging the two Fe atoms (b), and on the upper Fe atom (c). Panel (d) shows the total DOS for structure 5(c).